\documentclass[11pt,a4paper]{elsarticle}
\usepackage[utf8]{inputenc}
\usepackage{amsmath}

\title{Characterization of a fiber laser hydrophone\\ for acoustic neutrino detection}
\author[label1,label2]{E.~J.~Buis\corref{cor}}
\author[label1]{A.~M.~von~Benda-Beckmann}
\author[label1]{E.~Doppenberg}
\author[label1]{J.~Dorant}
\author[label1]{T.H.~Jansen}
\author[label1]{P.~Toet}
\author[label1]{P.~Verhooren}
\author[label1]{J.~de~Vreugd}
\cortext[cor]{ernst-jan.buis@tno.nl}
\address[label1]{TNO, Netherlands Organisation for Applied Sciences, Delft, The Netherlands}
\address[label2]{Nikhef, National Institute for Subatomic Physics, Amsterdam, The Netherlands}
\date{\today}

\begin{document}
\begin{abstract}
  This paper presents the development and characterization of a fiber laser hydrophone designed for deep-sea applications, with a focus on detecting neutrino interactions via their acoustic signatures. The hydrophone design includes a static pressure compensation mechanism, ensuring reliable operation at depths exceeding 1 km. The performance of the hydrophone was evaluated through laboratory tests and experiments in an anechoic basin, where its transfer function was measured before and after a 140-bar pressure cycle. The results show that the hydrophone maintains its sensitivity, with resonance peaks identified in both low- and high-frequency ranges. The hydrophone's sensitivity to acoustic signals was also compared to ambient sea state noise levels, demonstrating compatibility with the lowest noise conditions.
\end{abstract}
\maketitle

\section{Introduction}
In the search for ultra-high-energy neutrinos alternative detection techniques are being pursued. One of them is the detection of neutrino interactions through their acoustic signature in the deep sea \cite{Askaryan1957, Askaryan1979, Learned1979}. This method benefits from the relatively long attenuation length of the sound in water, allowing the establishment of an acoustic neutrino telescope with a large detection volume with a relatively sparse network of hydrophones. Still, for an acoustic telescope to observe the small predicted flux of ultra-high-energy neutrinos, the hydrophones are required to be sensitive enough to detect the weak signals induced by neutrino interactions. Additionally, achieving an effective detection volume of approximately 100 km$^3$ or larger would necessitate a large-scale array, comprising several thousand hydrophones \cite{Karg2006}.

Given the large number of hydrophones required for an acoustic neutrino telescope in combination with the required sensitivity, a promising technology is the fiber laser hydrophone technology as it combines sensitivity with a relatively simple and cost-effective method of deployment. Fiber optic hydrophones have been discussed in detail in literature, see e.g. \cite{Foster2005, Goodman2008}. For the application in an acoustic neutrino telescope, additional requirements have to be met on the performance of the hydrophones, such as the operation in the deep-sea environment ($>1$ km deep), sensitivity (comparable to the lowest sea state noise levels, SS0-SS1) and a large operating frequency bandwidth ($>10$ kHz).
The expected acoustic signals depend on neutrino energy but generally remain just above the sea state noise. For neutrino energies between \(10^{10}\) and \(10^{12}\) GeV, the signal amplitudes range from approximately 10 to several hundred millipascals at a distance of 1 km \cite{Bevan2007, Clara2020}.

In this paper we discuss the design of a new version of a hydrophone transducers that follows the design concepts based on the fiber laser technology as we have described previously \cite{Buis2013, Buis2017}; this time we focus on introducing a static pressure compensation mechanism. We present the results of an extensive set of measurements to characterize the new hydrophone transducer and to qualify for operation in the deep sea. The outline of this paper is as follows. In the next section we discuss the fiber hydrophone technology and the design of the transducer in particular. Then in section \ref{sec:exp} we detail the experimental set-up after which we show the results of the characterization measurements. We summarize in section \ref{sec:concl}.

\section{Hydrophone concept design}
\label{sec:design}
The concept of fiber hydrophones relies on the measurement of a wavelength shift in an optical fiber that is induced by sound pressure in water. A pressure wave will be converted to a wavelength shift by a transducer that is attached to the fiber. The sensitivity of this hydrophone technology depends on both the transducer’s ability to efficiently convert pressure into strain in the optical fiber and the fiber’s ability to translate that strain into a wavelength shift. This shift is then measured using a custom designed interferometer, which is a key component of the readout system. In this paper, we thoroughly examine and discuss all three essential elements: the transducer, the optical fiber, and the read-out system (or interrogator).

\subsection{Transducer design}
\label{sec:hydrophone_design}
\begin{figure}[ht]
 \begin{minipage}[]{0.495\textwidth}
   \begin{center}
   \includegraphics[width=0.95\textwidth]{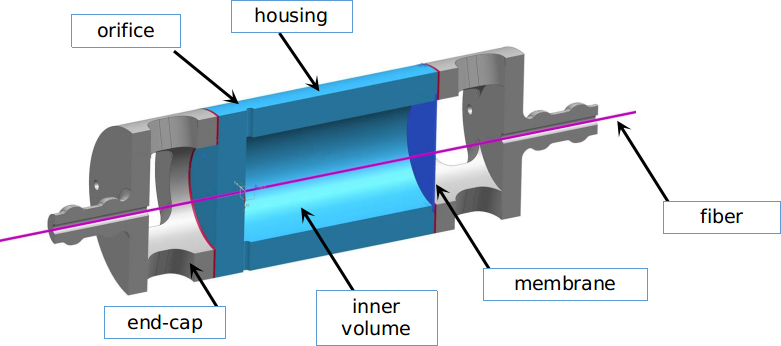}
   \end{center}
  \end{minipage}
  \begin{minipage}[]{0.495\textwidth}
   \begin{center}
  \includegraphics[width=0.95\textwidth]{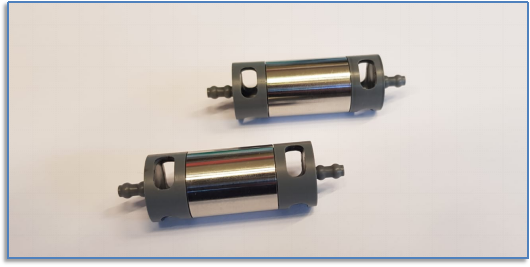}
   \end{center}
  \end{minipage}\\
  \begin{minipage}[]{0.495\textwidth}
    \begin{center}
      (a)
    \end{center}
  \end{minipage}
\hfill
  \begin{minipage}[]{0.495\textwidth}
    \begin{center}
      (b)
    \end{center}
  \end{minipage}
  \begin{center}
    \begin{minipage}[b]{\textwidth}
      \caption[]{\label{fig:sensor_design} (a) The design of the fiber laser hydrophone, while in (b) a photo of two assembled hydrophones are shown. In (a) the concept of a single membrane hydrophone is shown. }
  \end{minipage}
 \end{center}
\end{figure}

The transducer is in-house designed and is shown in figure \ref{fig:sensor_design}. The design relies on one (or two) membranes placed on a cylindrical shaped housing. The optical fiber is threaded through the center of the membranes and subsequently glued. The dimensions of the membrane, i.e., a thickness of 250 $\mu$m and a radius of 5 mm, were determined using COMSOL and define the  mechanical response of the transducer. The eigenfrequency, which corresponds to the position of the mechanical resonance peak, is particularly important, as the transducer's sensitivity is enhanced at this peak and declines sharply at higher frequencies.

The length of the transducer, i.e. 22 mm, is chosen to match  the length of the fiber laser cavity (next section). The transducer is made of stainless steel, which is chosen because of its stiffness, density and ease of manufacturing. Then, two versions of the transducers have been produced: i) one membrane on one side of the cylinder as depicted in figure \ref{fig:sensor_design} and ii) with a membrane on each side on the cylinder. Fabrication and integration of the transducer was done at TNO. The membrane was laser-beam welded (LBW) on to the cylinder, while the orifices were EDM-drilled. The fiber was integrated in the transducer and glued on the membrane with glue (Loctite\textsuperscript{\textregistered} EA9313). The fiber was not stripped from its coating before gluing. The pre-tension that was applied on the fiber amounted to about 0.1 N. Notably, the strength of the glue remained unchanged over time. After more than three years in storage, we remeasured the hydrophone transducer's response and found no detectable difference. Finally, to increase the robustness two end-caps are glued on either side of the cylinder to guide the fiber and to reduce the chance of breaking it. 

Particularly for this transducer, a hydrostatic pressure mechanism was introduced by an orifice in the housing, because the hydrophone is required to operate in the deep-sea and therefore to withstand high ambient pressure. The orifice allows water or oil to flow into the transducer to counterbalance the high pressure environment. Note however that a hydrophone with hydrostatic pressure compensation is expected to be less sensitive than an air-backed hydrophone. However, for an application in the deep-sea an air-backed volumes does not provide a solution, because the air is (completely) compressed by the ambient pressure.

\subsection{Fiber laser hydrophone concept and signal formation}
\begin{figure}[ht]
   \begin{center}
   \includegraphics[width=0.5\textwidth]{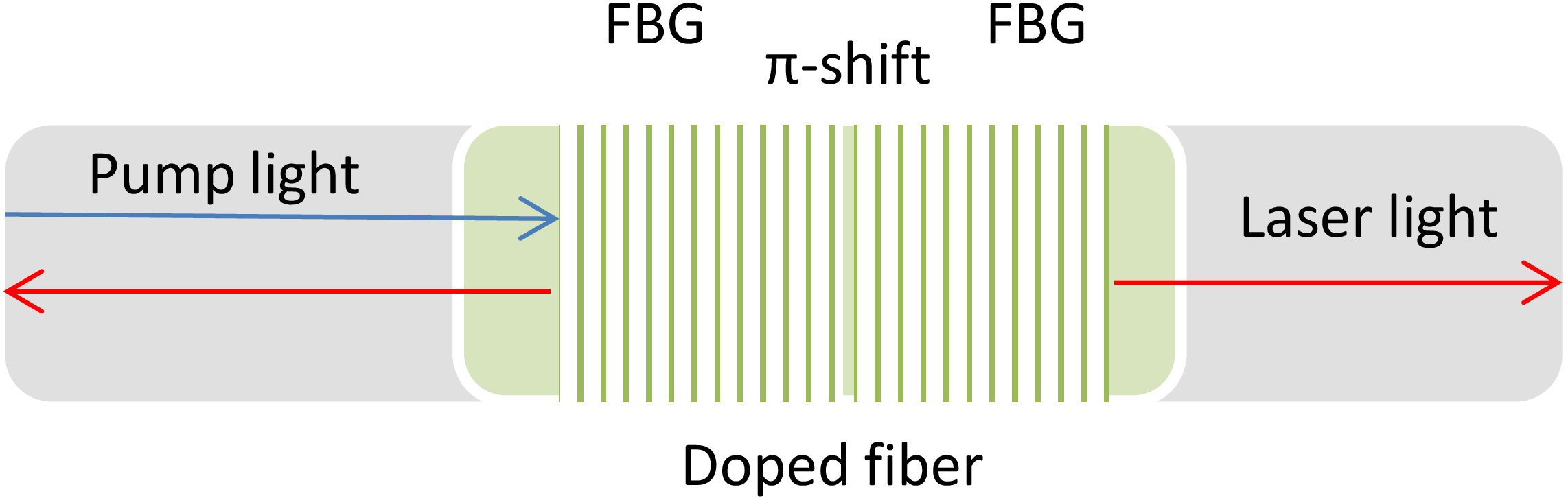}
   \end{center}
  \begin{center}
      \caption[]{\label{fig:fiberlaser} Schematic of a (distributed feedback) fiber laser. }
 \end{center}
\end{figure}

Dedicated optical fibers are realized that include a fiber laser as depicted in figure \ref{fig:fiberlaser} \cite{Cranch2008, Foster2017}. The gain medium of the fiber laser is provided by the Erbium-doped fiber core, while the fiber Bragg grating (FBG) confines the laser cavity. 

The concept of a fiber laser hydrophone is based on the fact that deformation (or strain) of the fiber laser cavity will result in a shift of the emitted laser wavelength. This change of wavelength can then be sensed using an interrogator, which is connected to the optical fiber. In order to increase the sensitivity of the hydrophones a transducer is placed around the fiber laser to convert pressure in strain in the laser cavity. An important parameter in the design of a transducer is the {\it strain gauge factor}, which is the amount of pressure that is converted to strain ($\varepsilon$) and is usually given in units n$\varepsilon$/Pa. The fiber laser used is a 22 mm long Erbium-doped Distributed Feedback Laser (DFB) manufactured by Exail. The fiber laser's length is a critical factor, as a shorter length maximizes strain for a given pressure wave. Additionally, the transducer housing is designed to match the length of the fiber laser.

For the discussion on the strain gauge factor it is illustrative to discuss the strain dependence of the Bragg wavelength. As mentioned above the strain ($\varepsilon$) in the fiber induces a wavelength shift ($\Delta\lambda_B$) which can be described by \cite{Kersey_1996}:
\begin{equation}
\label{eqn:strain_Kersey}
\Delta\lambda_B = (1 - p_e) \lambda_B \varepsilon
\end{equation}
where $\lambda_B$ is the central laser wavelength and $p_e$ is an effective photo-elastic coefficient:
\begin{equation}
p_e = \frac{n}{2}(P_{12} - \mu(P_{11} + P_{12}))
\end{equation}
Here $P_{ij}$ are the Pockels coefficients of the optical fiber, $n$ is the index of refraction and $\mu$ is the Poisson's ratio \cite{Bertholds88}. Given the values of $p_e \sim 0.22 $ and $\lambda_B \sim 1550\;{\rm nm}$, the wavelength shift dependence on the strain is given by:
\begin{equation} 
\label{eqn:dlambda_depsilon}
\frac{\Delta \lambda_B}{\varepsilon} = 1.2 \left( \frac{pm}{\mu\varepsilon} \right)
\end{equation} 

As we will discuss in section \ref{sec:exp}, the hydrophone is read out using an unequal-arm interferometer in which phase difference between light from the different arms is determined by introducing an optical path difference (OPD) in two arms. This phase difference ($\phi$) is defined as: 
\begin{equation}
\label{eqn:phase}
\phi = \frac{2\pi}{\lambda_B}\textsc{OPD}
\end{equation}
A change in the central wavelength of the fiber laser induces a phase difference, which depends on the central wavelength and is given (for small wavelength changes) by:
\begin{equation}
\label{eqn:dphase_dlambda}
    \frac{\Delta\phi}{\Delta\lambda_B} = \frac{2\pi}{\lambda_B^2}\textsc{OPD}
\end{equation}

The expressions as given in equations \ref{eqn:dphase_dlambda} and \ref{eqn:dlambda_depsilon} are used to convert the measured phase to the strain in the fiber laser. The value OPD should be corrected for the index of refraction, i.e. the OPD is this case is $n$ times larger than the physical length of the difference in the arm length, where $n$ is the index of refraction of the glass fiber. In principle a larger $\textsc{OPD}$ provides a higher sensitivity, but large arm lengths in the interferometer are unpractical and susceptible for additional (internal fiber) noise. It was found that in our experimental set-up the optimal optical path difference was between 4 and 10 meter.

\section{Experimental setup}
\label{sec:exp}

\subsection{Interrogator}
\begin{figure}[ht]
   \begin{center}
   \includegraphics[width=0.75\textwidth]{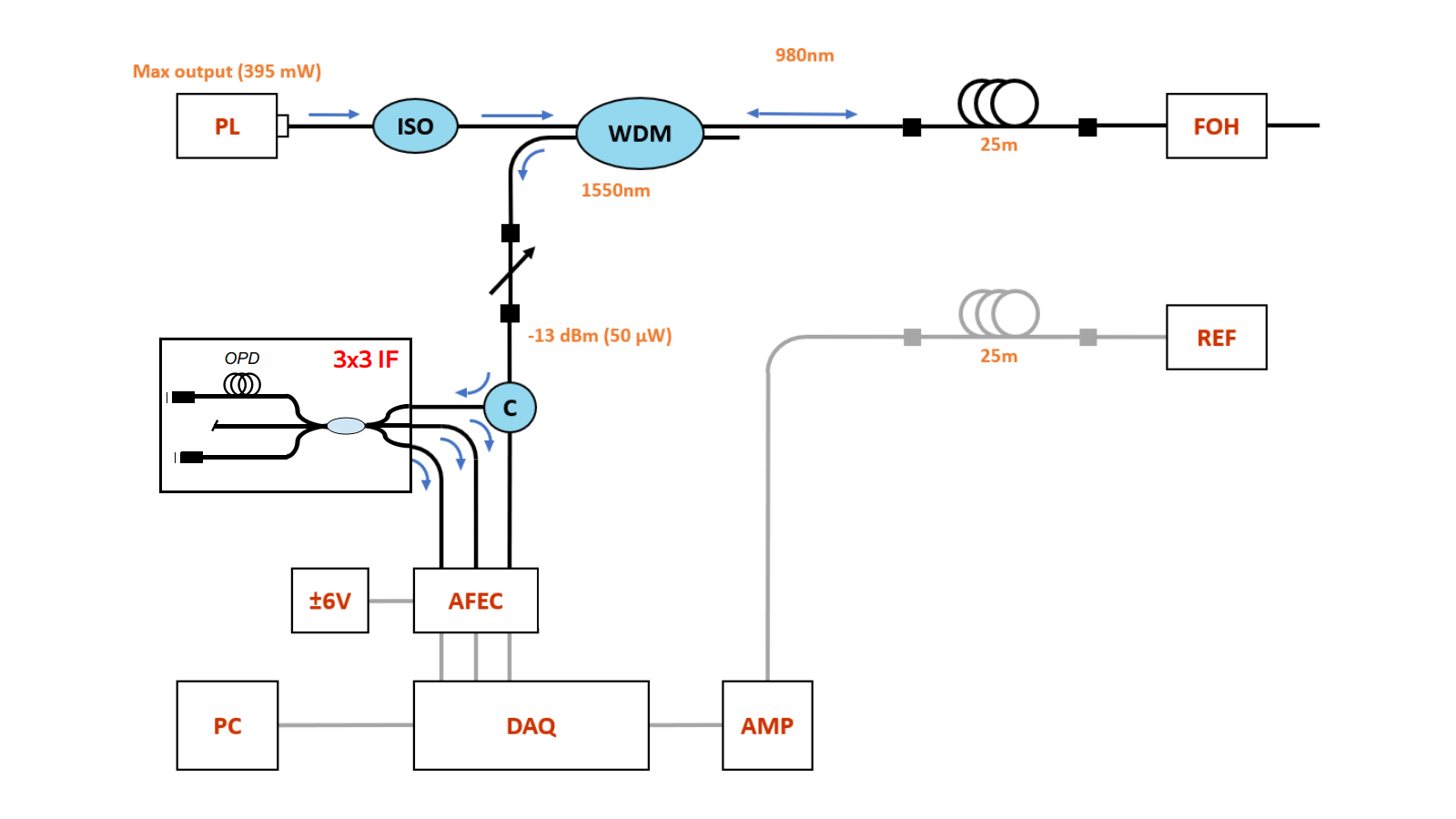}
   \end{center}
  \begin{center}
      \caption[]{\label{fig:interrogator} Schematics of the read out of a fiber laser hydrophone. Several abbreviations are used in the drawing: PL = pump laser; ISO = isolator; WDM = wavelength division multiplexer, FOH= fiber optic hydrophone; REF = reference hydrophone; IF = interferometer; C = circulator; AMP = amplifier; AFEC = Analog Front-End Controller  }
 \end{center}
\end{figure}

The read-out system of the fiber hydrophone is shown in figure \ref{fig:interrogator} and consist of three main components. First, a pump laser is used to pump the fiber laser. The pump laser wavelength can be either 980 or 1480 nm. Here we use 980 nm. 

Secondly, an interferometer is used to determine the wavelength shift of the fiber laser light. We use a custom built interferometer for this based on a standard 3x3 fiber coupler. While the input of the interferometer consists of one input and three outputs, we only used two arms in the interferometer. A fixed and predetermined optical path difference (OPD) in the two arms has been chosen.
The output of the 3x3 coupler consists of three light intensities ($I_x$) as indicated in figure \ref{fig:interrogator}. The outputs of the three arms are identical, but each is shifted in phase over $\frac{2}{3}\pi$. In addition, each intensity depends on the optical path difference as defined in the equation \ref{eqn:3_intensities}.
\begin{eqnarray}
  \label{eqn:3_intensities}
I_S &=& I_0(1+V\cos(\tfrac{2\pi}{\lambda}\textsc{OPD}))  \nonumber \\
I_+ &=& I_0(1+V\cos(\tfrac{2\pi}{\lambda}\textsc{OPD} + \tfrac{2\pi}{3})) \nonumber \\
I_- &=& I_0(1+V\cos(\tfrac{2\pi}{\lambda}\textsc{OPD} - \tfrac{2\pi}{3}))
\end{eqnarray}
where $I_0$ is the normalized amplitude of the signal and $V$ is the visibility. 
Using the phase shift and the intensity of each arm, the optical path length ($\textsc{OPD}$) is given by the following equation:  
\begin{eqnarray}
  \label{eqn:OPD}
  \phi &=& \arctan\left(\tfrac{\sqrt{3} (I_+ - I_-)}{2I_s - I_+ - I_-} \right) \\
  \textsc{OPD} &=& \tfrac{\lambda}{2\pi} \arctan\left(\tfrac{\sqrt{3} (I_+ - I_-)}{2I_s - I_+ - I_-} \right) 
\end{eqnarray}
Because the optical path difference ($\textsc{OPD}$) is known, the wavelength shift can be accurately determined when the three output intensities are measured. 

Then, thirdly, we convert the three optical signals from the interferometer arms into a digital signal. To achieve this, we collect the light intensities from the interferometer using commercially available photo-diodes, which are then converted using a 24-bit ADC at a sample rate of 200 kHz. Finally, we use the same DAQ system to register the signals from a commercially available reference hydrophone (section \ref{sec:acoustic_setup}).

\subsection{Phase measurement calibration}
\begin{figure}[ht]
 \begin{minipage}[]{0.495\textwidth}
   \begin{center}
   \includegraphics[width=0.95\textwidth]{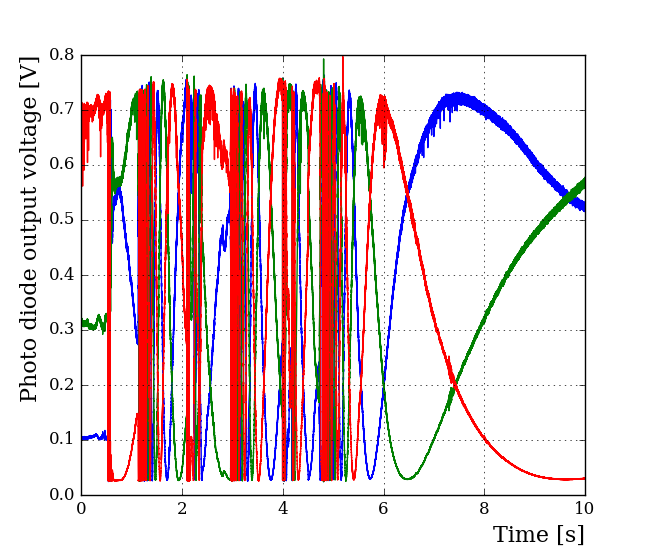}
   \end{center}
  \end{minipage}
  \begin{minipage}[]{0.495\textwidth}
   \begin{center}
  \includegraphics[width=0.95\textwidth]{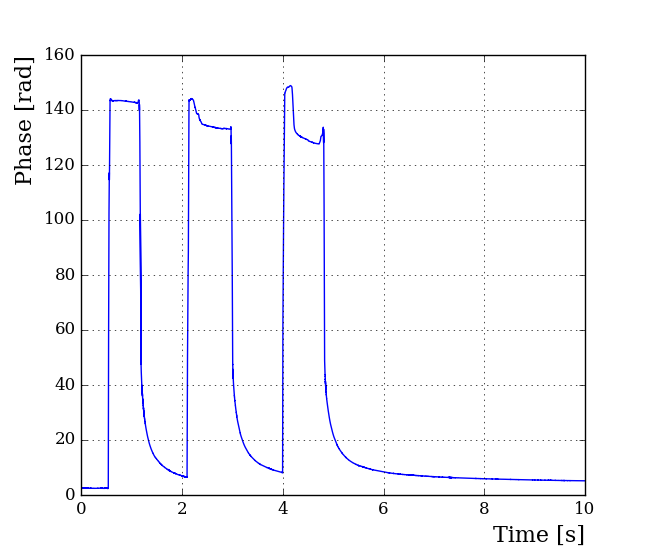}
   \end{center}
  \end{minipage}\\
  \begin{minipage}[]{0.495\textwidth}
    \begin{center}
      (a)
    \end{center}
  \end{minipage}
\hfill
  \begin{minipage}[]{0.495\textwidth}
    \begin{center}
      (b)
    \end{center}
  \end{minipage}
  \begin{center}
    \begin{minipage}[b]{\textwidth}
      \caption[]{\label{fig:phase_calibration} (a) The output of the three arms as a function of time. At three moments in time, strain is applied on of the arms in the interferometer, large enough to induce a phase shift of more than 2$\pi$ radians. In (b) the resulting phase shift is shown as a function of time. }
  \end{minipage}
 \end{center}
\end{figure}

As discussed above the transducer will convert pressure waves into strain, resulting in a wavelength shift emitted by the fiber laser. We use a 3x3 interferometer to determine the wavelength shift, by measuring the phase shift that the change of wavelength introduces. 
To accurately determine the phase shift, we first calibrate the output of the three arms of the interferometer. In figure \ref{fig:phase_calibration} (a) we show the output of the three interferometer arms as a function of time. At certain moments in time we apply strain on the one the arms (by stretching the fiber by hand), which is large enough to exert phase shifts larger than 2$\pi$ radians. From these fringes, we can then correct the gain and phase of each arms in the offline analysis. This achieved by calculating the gain correction factor that has to be applied to each of the three outputs in figure \ref{fig:phase_calibration} to have a normalized value. In addition a phase correction can be found by requiring that the phases differ by  $\frac{2}{3}\pi$. When the gain and phase correction is applied, we obtain the phase as a function of time according equation \ref{eqn:OPD} and which is shown in figure \ref{fig:phase_calibration} (b). From this figure is clear that strain was induced at 0.5, 2 and 4 seconds after the start of the measurement. 

\subsection{Acoustic setup}
\label{sec:acoustic_setup}
\begin{figure}[ht]
   \begin{center}
   \includegraphics[width=0.75\textwidth]{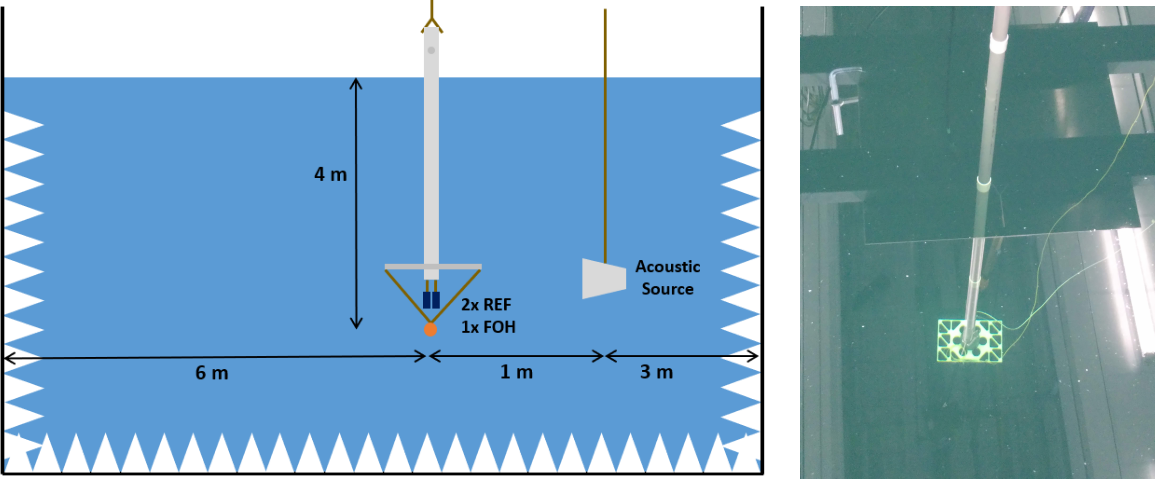}
   \end{center}
  \begin{center}
      \caption[]{\label{fig:acoustic_setup} The left panel shows a schematic of the acoustic measurement setup in the anechoic basin. The right panel displays a photograph of the mechanical support structure, to which both the fiber hydrophone and the reference hydrophone are attached.}
 \end{center}
\end{figure}

To characterize the fiber hydrophone, the experimental setup as described above was taken to the anechoic basin at the TNO laboratories. The experimental set-up in the basin is depicted schematically in figure \ref{fig:acoustic_setup}. The basin itself measures 8x8x10 m$^3$.
Two projectors were used: i) an ACTRAN projector (type LFPX-4), which provides acoustic signals for the lower frequencies and ii) an ITC 1042 projector which provides an acoustic source at higher frequencies, i.e. above 10 kHz.

A mechanical support structure was produced to which both the fiber hydrophone and the reference hydrophone were attached.  

\subsection{Pressure calibration}
\label{sec:calibration}
The calibration of the fiber hydrophone is established using the reference hydrophone, i.e a commercial available Bruel \& Kjaer BK8103. In turn this hydrophone is calibrated using a pistonphone (type BK4229 with coupler UA 0548), which provides a well-defined signal at 250 Hz to the coupled reference hydrophone. This signal has been calibrated and provides a well defined signal of $166\;{\rm dB\;re}\;\mu{\rm Pa}$ to the reference hydrophone. The response of the BK8103 is specified to be flat up to 50 kHz, so that the calibration of the reference hydrophone is valid for the range at which the fiber hydrophone is tested. Both the reference hydrophone and the pistonphone undergo periodic control measurements. Moreover, cross measurements with various reference hydrophones have been carried out for additional verification of the hydrophone's response.

\section{Measurement results}
\label{sec:results}
\subsection{Hydrophones under test}
To assess the performance of the fiber hydrophones a number of different configurations have been integrated and characterized. In table \ref{tab:hydrophones_under_test} the various fiber hydrophone configurations are listed. Each of the fiber hydrophones have been characterized, but in addition we performed a number of measurements to better understand the impact of several issues during integration. One fiber hydrophone has been subjected to a pressure qualification measurement.
\begin{table}[ht]
    \centering
    \begin{tabular}{|l|l|}
    \hline
    Configuration & Remarks \\ \hline
    Single membrane & with and without cap\\
    Single membrane & with and without air bubble \\
    Double membrane & before and after pressure cycle\\
    \hline
    \end{tabular}
    \caption{Devices integrated for characterization.}
    \label{tab:hydrophones_under_test}
\end{table}

All fiber hydrophones were filled with water. To properly fill the volumes within the transducers, an experimental set up was made to fill the transducers without air bubbles. After integration the transducers were put in a reservoir under vacuum and water was led in the reservoir. This controlled way of filling ensured that the amount of air in the volumes after being filled was minimal.  

\subsection{Noise}
\begin{figure}[ht]
   \begin{center}
   \includegraphics[width=0.5\textwidth]{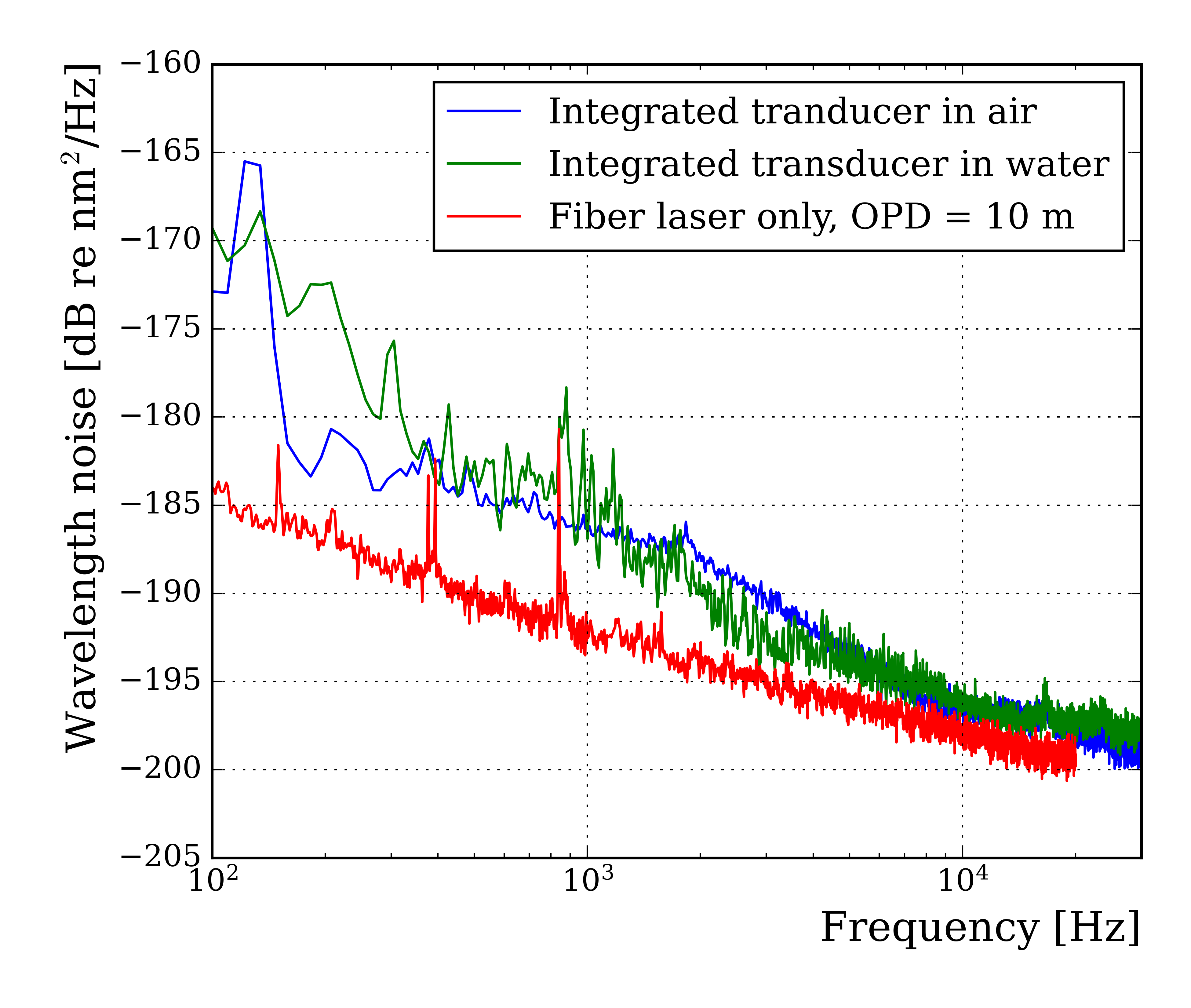}
   \end{center}
  \begin{center}
      \caption[]{\label{fig:RIN_fl} Power spectral density of the phase noise of a fiber laser integrated in a fiber hydrophone compared to the noise of an acoustically insulated bare fiber laser.}
 \end{center}
\end{figure}

Prior to integrating the fiber hydrophones, bare fiber lasers coupled to the interrogator were characterized. Using the interrogator the phase noise of the fiber laser without an attached transducer was measured. The fiber laser was kept in an acoustically isolated block of aluminium. In  addition the interrogator was kept free from vibrations. The set-up used included an interferometer with a optical path length (OPD) of 10 m. Hence, the measurements on the bare fiber laser presents the noise floor of the interrogator. In figure \ref{fig:RIN_fl} the wavelength noise - obtained from the phase shift using equation \ref{eqn:dphase_dlambda} - for one of the bare fiber lasers is shown for as a function of the frequency. For comparison we show the noise measurement on a packaged hydrophone both in water and in air. Here a  set-up that included an interferometer with an OPD of 4 m as used. As can be seen from the figure, the noise level of the integrated and the bare fibers do not differ much at higher frequencies, i.e. above 1 kHz. At a frequency of 10 kHz, the difference in the noise level is about 2 dB. At lower frequencies, the integrated fiber picks up both acoustic and vibration noise. Peaks in the noise spectrum (clearly seen in the fiber laser spectrum) originate from electromagnetic interference, as some noise is coupled into the analog front-end of the interrogator.

\subsection{Time trace and Bode diagram}
\begin{figure}[ht]
   \begin{center}
   \includegraphics[width=0.65\textwidth]{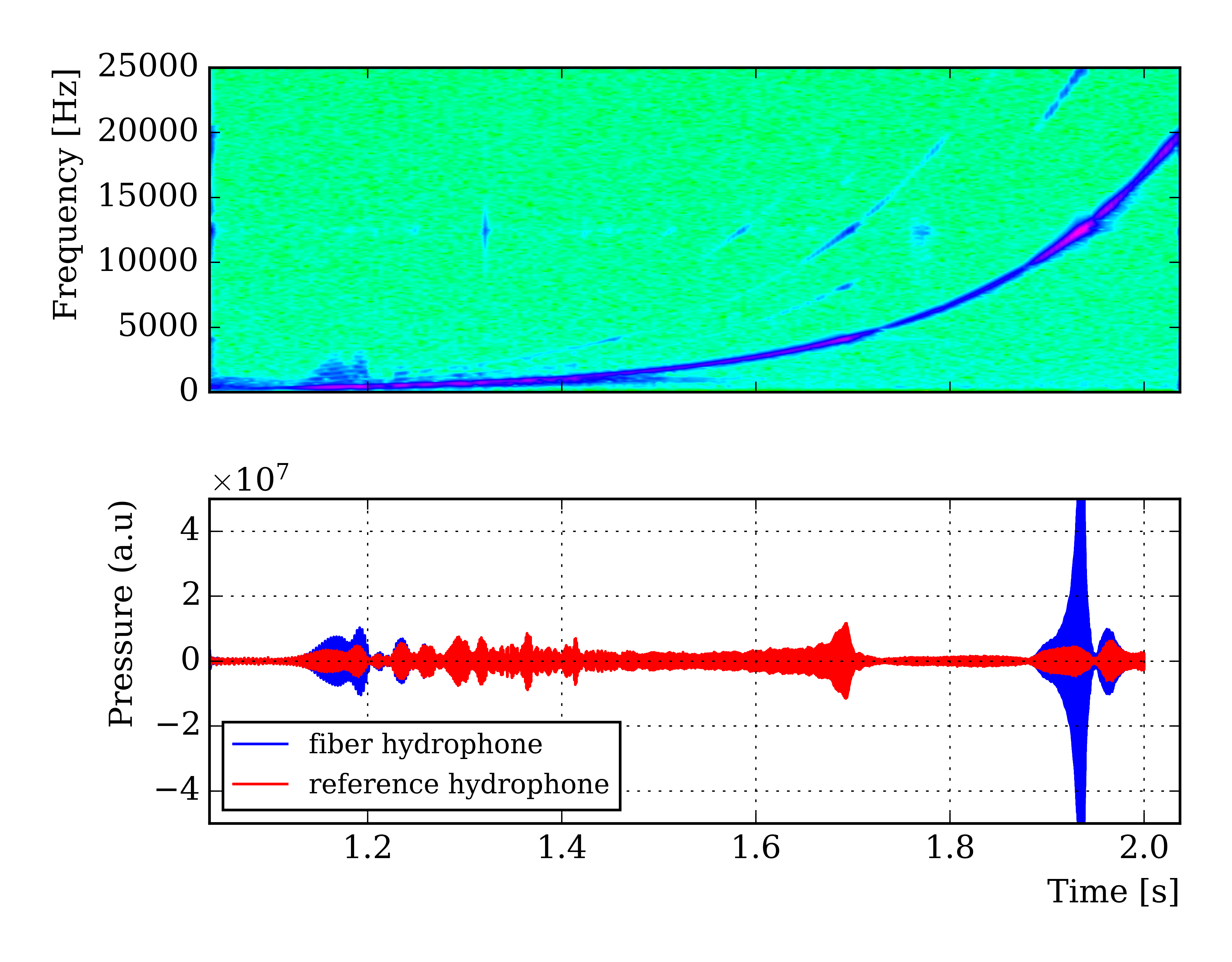}
   \end{center}
  \begin{center}
      \caption[]{\label{fig:spectrogram} Example spectrogram of a single membrane fiber hydrophone (top panel). The colors in this plot qualitatively represent the increasing signal amplitude, transitioning from green to purple. In the bottom panel the time traces from the fiber and reference hydrophones in a frequency sweep measurement are shown. }
 \end{center}
\end{figure}

The fiber hydrophone was calibrated and deployed in the acoustic basin as discussed in section \ref{sec:exp} and depicted in figure \ref{fig:acoustic_setup}. One of the basic measurements is a frequency sweep in which the output is measured from both the fiber hydrophone and the reference hydrophone. The result from one such sweep is plotted in figure \ref{fig:spectrogram}. The upper panel shows the spectrogram of the output of the fiber hydrophone. A clear increase of the acoustic frequency from 200 Hz to 20 kHz in the blue and purple band. In addition, the spectrogram shows some of the overtones at twice the nominal frequency in a dim blue band. It was found that these overtones originate mainly from harmonic distortions of the input amplifier of the projector.

In the lower panel of figure \ref{fig:spectrogram} the time trace for both the fiber hydrophone and the reference hydrophones are shown. Here, we present the uncalibrated data to demonstrate that (i) the output of the projector, as recorded by the reference, is not flat across the spectrum, and (ii) the transfer function of the fiber hydrophone is also not uniform, but rather complex. It shows the importance of the reference hydrophones in these measurements to correct for the distortions in the source and the investigate the transfer function of the fiber hydrophone.

Frequency sweeps were used to determine transfer functions and were plotted in general Bode diagrams consisting of a Bode magnitude plot and a Bode phase plot. In addition we determine the coherence between the signals in the fiber hydrophone and the reference hydrophone. The coherence confirms whether the measured signal is observed in both hydrophones and hence whether the input is of acoustic origin. We compute (magnitude of the) transfer function ($T_{xy}$) and the coherence ($C_{xy}$) according:
 \begin{eqnarray}
T_{xy}(f) &=& \frac{|S_{xy}(f)|}{S_{xx}(f)}   \\ 
C_{xy}(f) &=& \frac{|S_{xy}(f)|^2}{S_{xx}(f)S_{yy}(f)}    
\end{eqnarray}
where $S_{xy}(f)$ are the cross-spectral density between fiber hydrophone and the reference hydrophone, and $S_{xx}(f)$ and $S_{yy}(f)$ the auto spectral density of fiber and reference hydrophone respectively. Furthermore, the phase in the Bode diagram is defined as the angle between input of at the reference hydrophone and fiber hydrophone.

 \begin{figure}[ht]
   \begin{center}
   \includegraphics[width=0.65\textwidth]{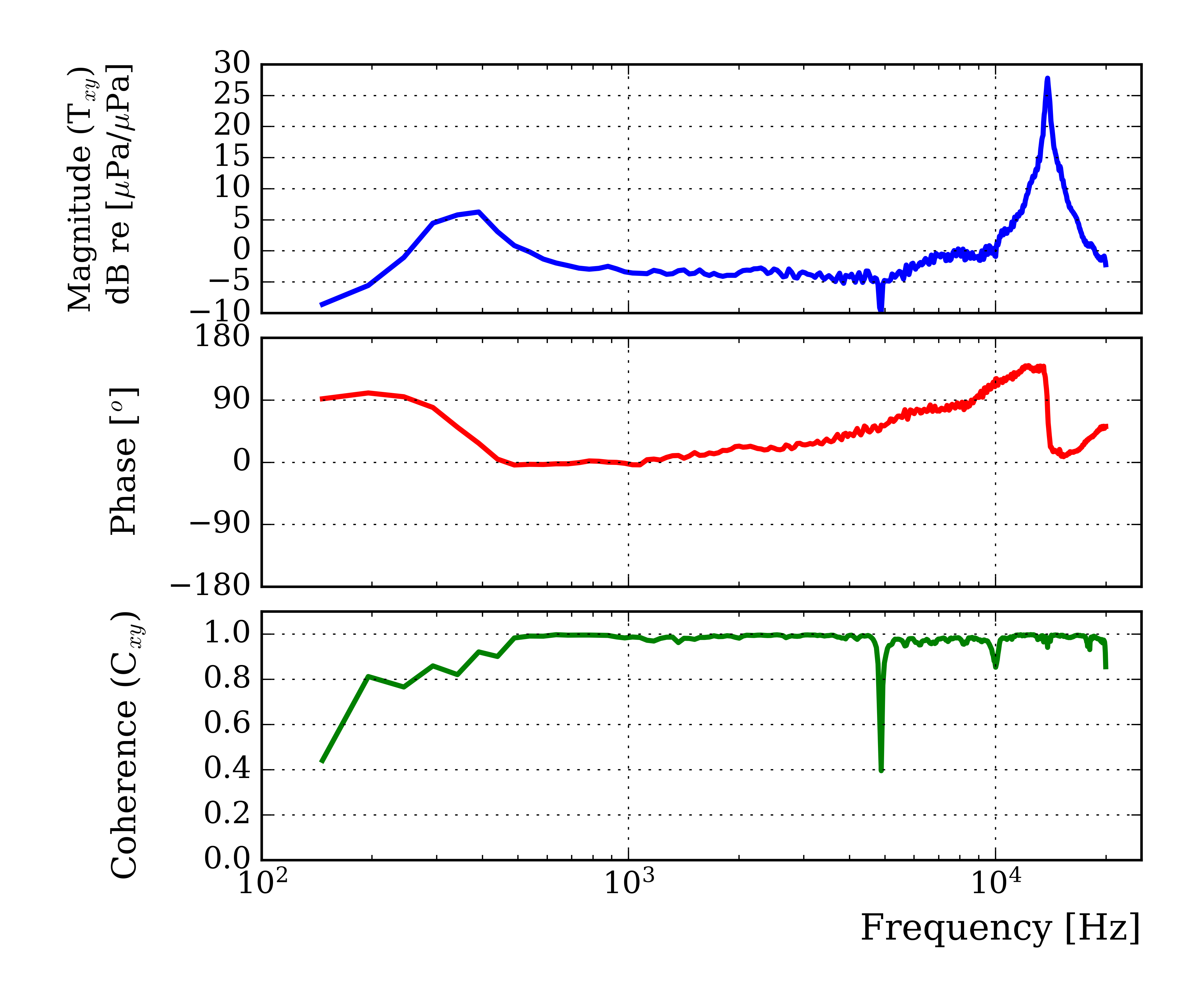}
   \end{center}
  \begin{center}
      \caption[]{\label{fig:bodeplot} Response of a single membrane hydrophone to a sweep signal with a frequency between 200 Hz and 20 kHz. In the top panel the Bode magnitude plot is shown; in the middle panel the Bode phase plot is shown, while the bottom panel shows the coherence of the signal between the fiber hydrophone and the reference hydrophone.}
 \end{center}
\end{figure}
In figure \ref{fig:bodeplot} we show the Bode plot of a frequency response measurements in which an acoustic frequency sweep between 200 Hz and 20 kHz was offered to both the (single membrane) fiber hydrophone and the reference hydrophone. The panels in this figure show (from top to bottom) the magnitude and the phase of the transfer function as well as the coherence (cross spectral density) between the fiber hydrophone and the reference. The coherence in the frequency range of 200 Hz - 20 kHz is on average high, albeit that there are several dips, which can be attributed to non-acoustic signals such as electronic noises.

The transfer function shows several features as shown in figure \ref{fig:bodeplot}. At low frequencies, i.e. below $\sim$ 600 Hz a resonance peak can be found, which is a Helmholtz resonance due to the small orifices in the transducer. When the orifices were closed off using a piece of tape, the Helmholtz resonance disappeared. At higher frequencies, i.e. above 10 kHz a peak in the transfer function appears, which is the first eigenfrequency and is due to a mechanical resonance. At the position of the resonance peaks, a phase jump is shown in the Bode phase plot in the middle panel.

\begin{figure}[ht]
   \begin{center}
   \includegraphics[width=0.5\textwidth]{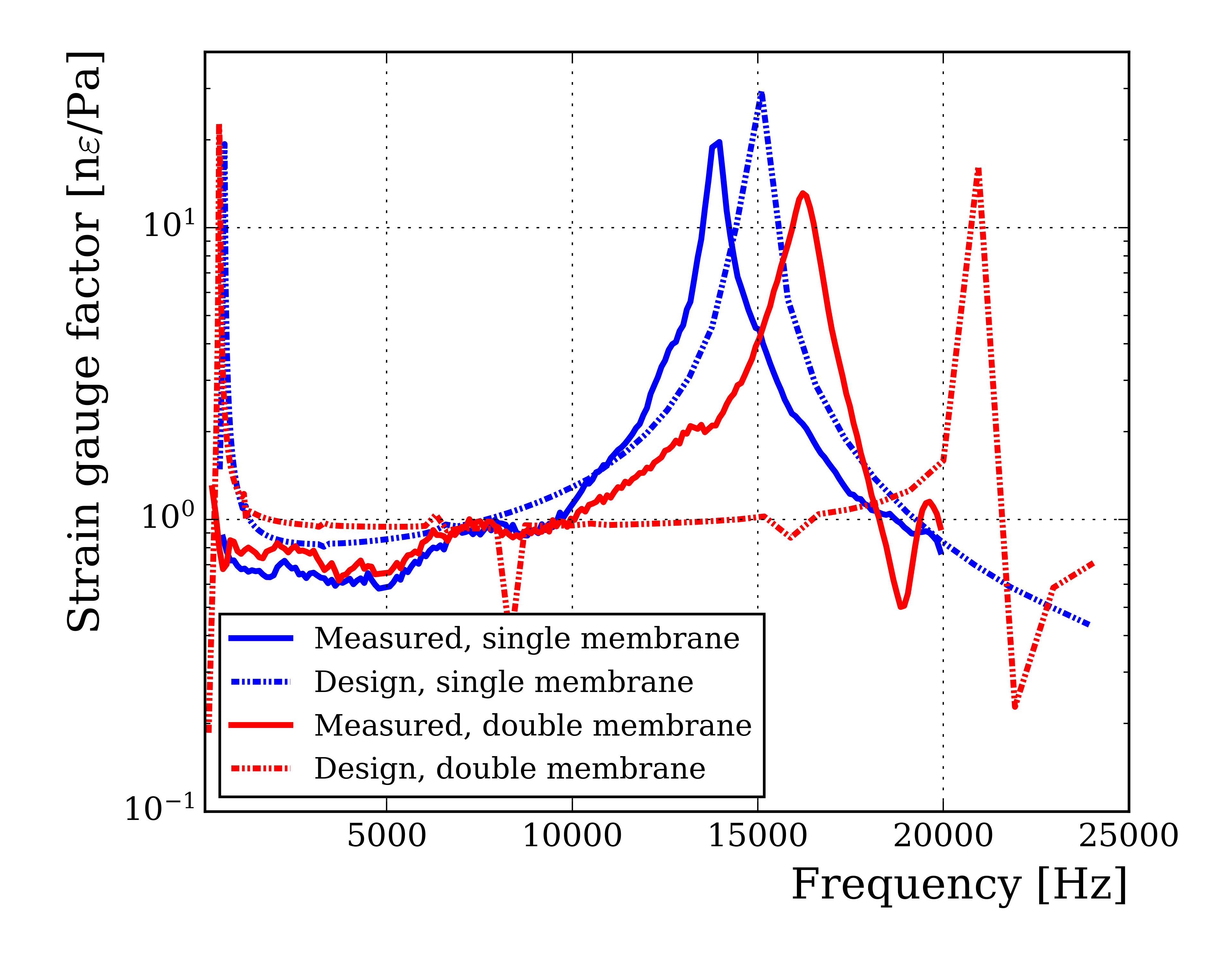}
   \end{center}
  \begin{center}
      \caption[]{\label{fig:strainsensitivity_design} Measured and designed strain gauge factor of a single and double membrane fiber hydrophone as a function of the frequency. }
 \end{center}
\end{figure}

Bode plots are used to determine the strain gauge factor of the transducer. From the measured laser phase shift (and hence wavelength shift) the strain gauge factor can be determined using equation \ref{eqn:dlambda_depsilon}. In figure \ref{fig:strainsensitivity_design} we show the strain gauge factor for both the single and double membrane fiber hydrophones. The gauge factor is given for the frequency range in which a frequency sweep has been carried out and in which the coherence is sufficient (i.e. $C_{xy} > 0.7$). As we characterized two single membrane transducers, we compared their transfer functions, and no significant differences were found. 

In addition, the strain gauge factor as predicted by the COMSOL finite element analysis is given. As can be seen from the figure, the measured strain gauge factor reasonably matches the valued determined using the design software, albeit that the position of the resonance peak is found at lower frequencies than originally designed.

\subsubsection{Impact of protective cap}
\begin{figure}[ht]
 \begin{minipage}[]{0.495\textwidth}
   \begin{center}
   \includegraphics[width=0.95\textwidth]{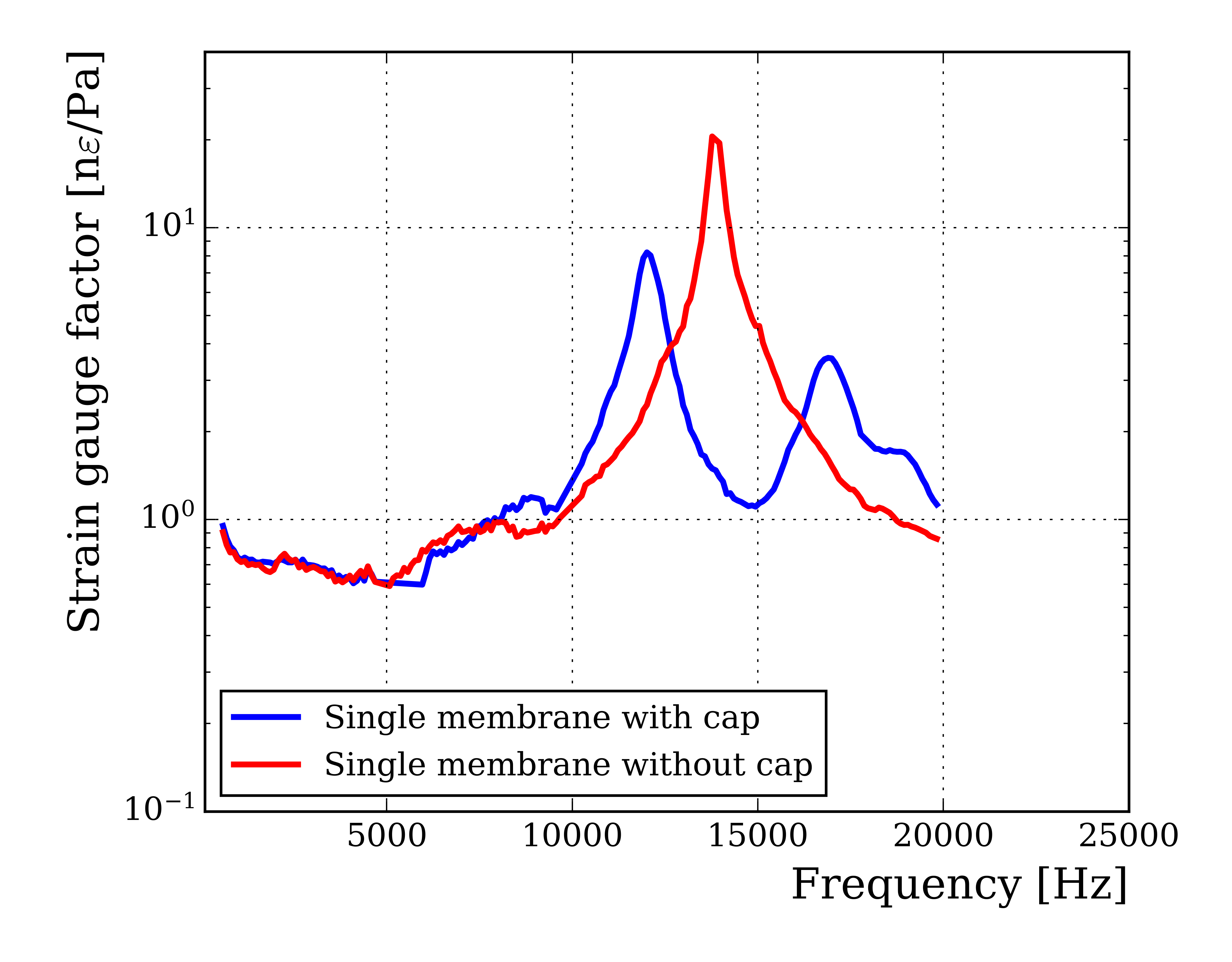}
   \end{center}
  \end{minipage}
  \begin{minipage}[]{0.495\textwidth}
   \begin{center}
  \includegraphics[width=0.95\textwidth]{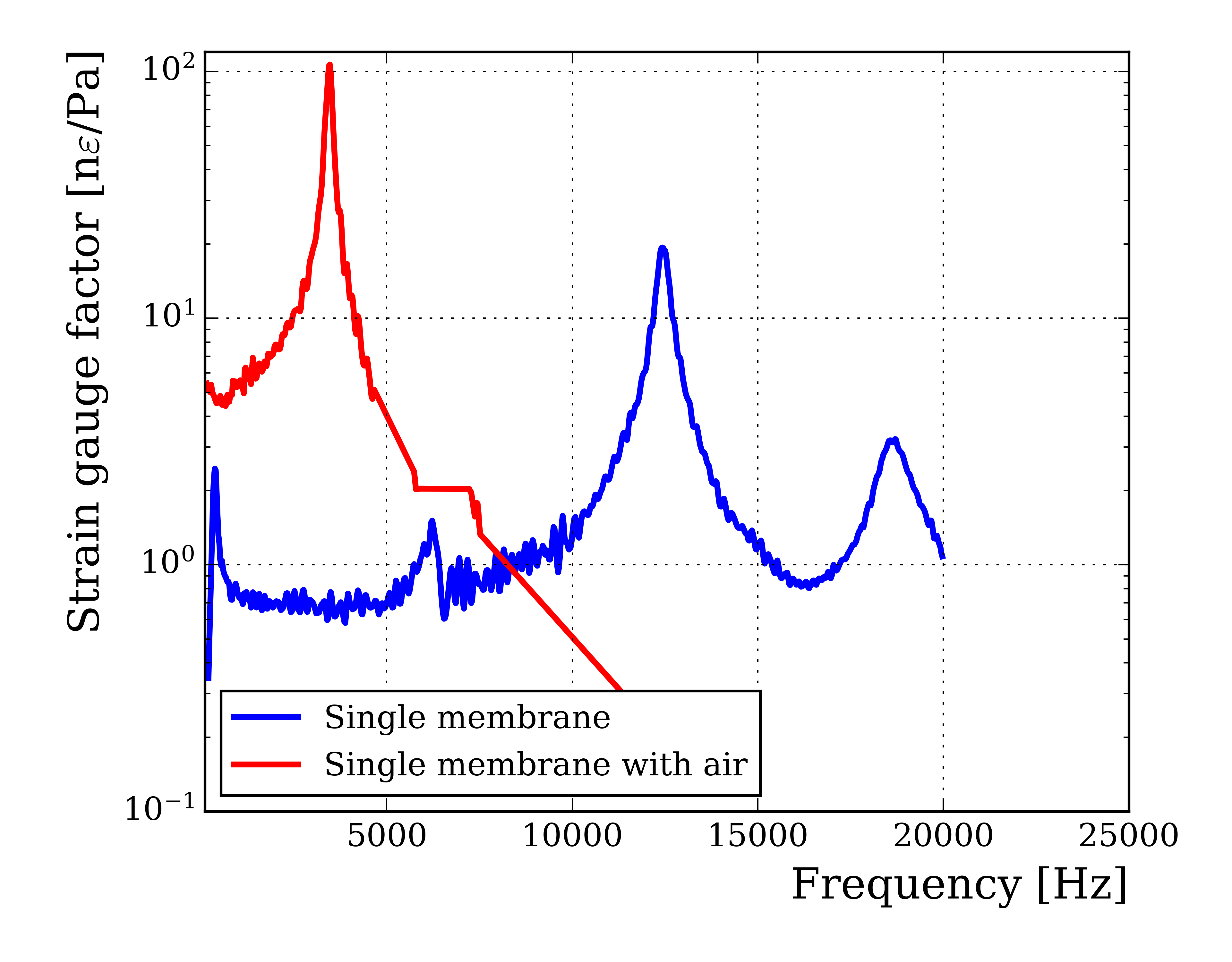}
   \end{center}
  \end{minipage}\\ 
  \begin{minipage}[]{0.495\textwidth}
    \begin{center}
      (a)
    \end{center}
  \end{minipage}
\hfill
  \begin{minipage}[]{0.495\textwidth}
    \begin{center}
      (b)
    \end{center}
  \end{minipage}
  \begin{center}
    \begin{minipage}[b]{\textwidth}
      \caption[]{\label{fig:cap_and_air} (a) The fiber hydrophone response as a function of the acoustic frequency for a single membrane transducer with a cap compared to a single membrane transducer without a protective cap. In (b) the transfer functions of a sensor with a sensor volume completely filled with water and with an air bubble inserted.}
  \end{minipage}
 \end{center}
\end{figure}

As shown in figure \ref{fig:sensor_design} a cap is glued on both ends of the transducer to guide the fiber through the membrane of the transducer. The cap was added to the transducer for practical reasons, to avoid breaking the optical fiber during a extensive set of measurements and handling. To determine the impact of the cap on the acoustic characteristics of the fiber hydrophone, the transfer function of the fiber hydrophone has been determined with and without the cap. In figure \ref{fig:cap_and_air} (a) the transfer function of the single membrane sensor is shown for both configurations. As can be seen from the figure, the cap has a non-negligible impact on the acoustic characteristics: the main mechanical resonance has shifted to a lower eigenfrequency. In addition, a second resonance peak can be seen that is absent when no cap has been applied. The main mechanical resonance peak shifts from 13.9 kHz to a frequency of 12.0 kHz. Furthermore there are two small anti-resonance peaks, at 4.8 and 5.7 kHz respectively. The latter is introduced by the cap and shows directionality, while the first anti-resonance peak is always present, irrespective of the cap or direction w.r.t to the sound source. The origin of the peak is unknown. Note further that the Helmholtz resonance peak remained at its position.

\subsubsection{Air bubbles in the sensor}
As mentioned in section \ref{sec:hydrophone_design}, a static pressure mechanism has been implemented by introducing two orifices in the sensor. Prior deployment the sensor volume has been filled with water, which allows to withstand the ambient pressure when the sensor is required to operate at large depth. Filling the sensor volume with water however, (instead of air) does compromise the sensitivity of the sensor. It is important to completely fill the sensor as any residual air will provide a risk to deform and damage the sensor. It is anticipated that residual air is expected to have an impact on the transfer function. To verify the impact of air in the sensor, we have determined the transfer function without air, after careful filling of the sensor and with air. Then, a small bubble of air was inserted in the sensor using an injection needle. Unfortunately we did not have full control of the size of air bubble, but we estimated its size to be small and around 1\% percent in volume only with respect to the sensor volume. 

In figure \ref{fig:cap_and_air} (b) we show the transfer function with and without air bubble in the sensor. Comparing the transfer functions differ both in the position of the mechanical resonant frequency as well as the overall strain gauge factor. After introducing air in the sensor volume, the peak shifts to lower frequencies and the overall sensitivity seems to increase. The shape of the transfer function did not change. The impact of the air bubble is confirmed by simulations.

\subsection{Linearity}
\begin{figure}[ht]
   \begin{center}
   \includegraphics[width=0.5\textwidth]{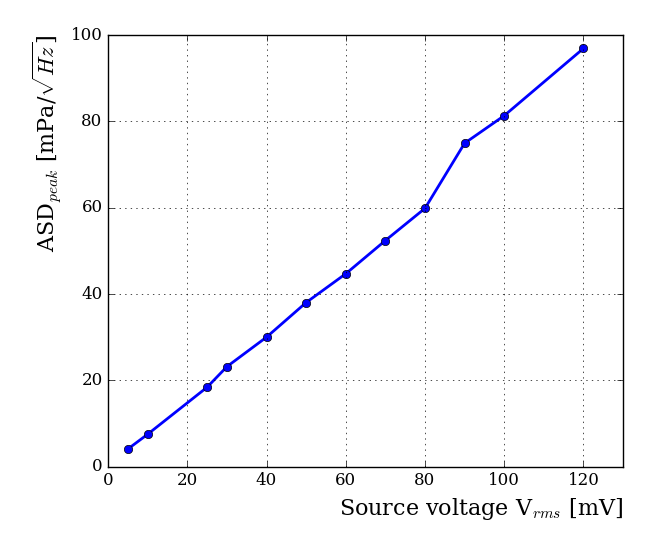}
   \end{center}
  \begin{center}
      \caption[]{\label{fig:linearity} Linearity of a single membrane sensor.}
 \end{center}
 \end{figure}

To check the linearity of the fiber hydrophone, its response to a tone of 5 kHz was determined at various sound source settings. The sound pressure level of the source was set by a voltage and the recorded sound pressure level was determined for a tone of 5 kHz. In figure \ref{fig:linearity} the recorded height of the sound pressure as a function of the voltage applied to the source is shown. The sensor shows a linear behavior in the applicable range. 

\subsection{Directionality}
\begin{figure}[ht]
 \begin{minipage}[]{0.495\textwidth}
   \begin{center}
   \includegraphics[width=0.95\textwidth]{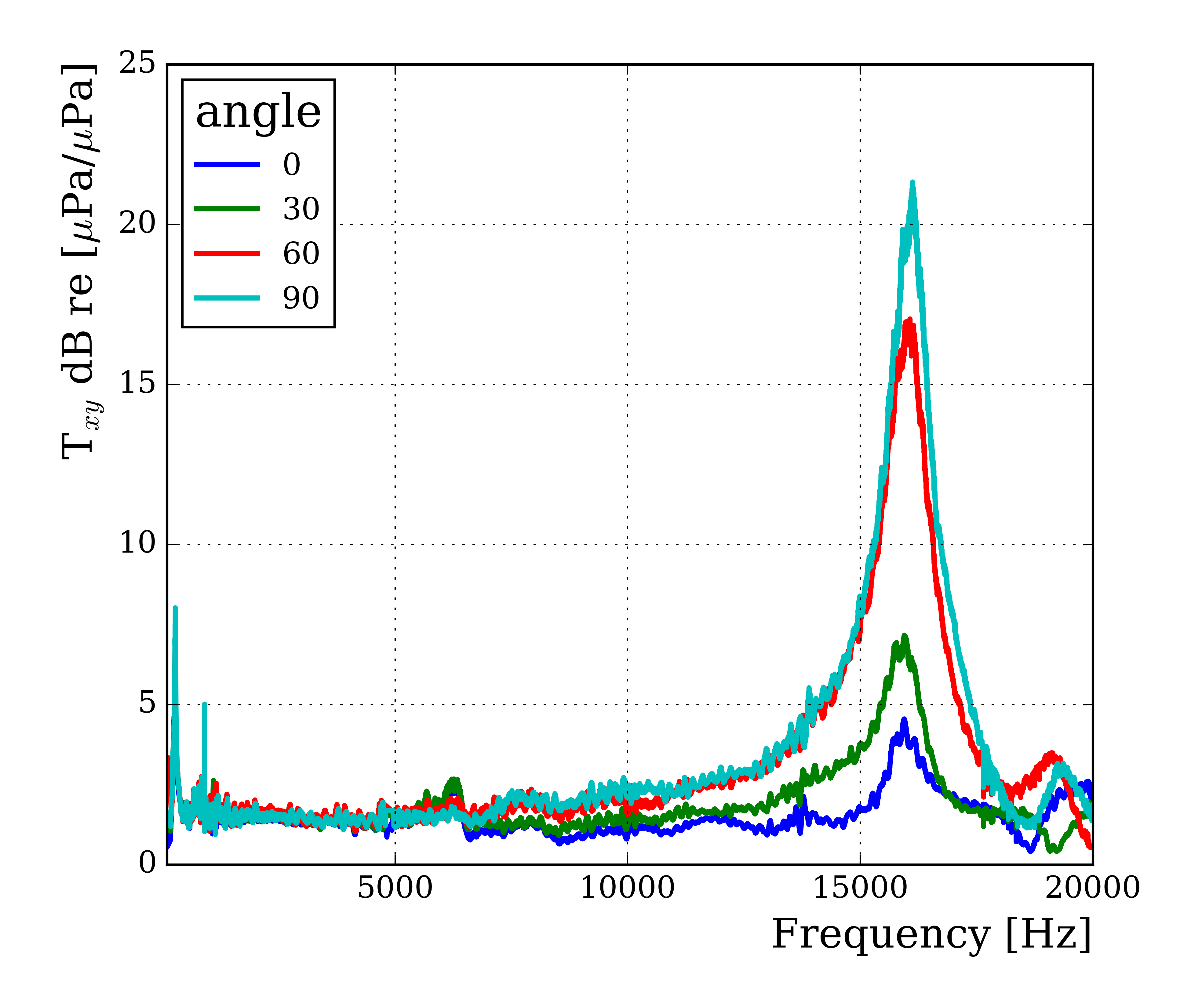}
   \end{center}
  \end{minipage}
  \begin{minipage}[]{0.495\textwidth}
   \begin{center}
  \includegraphics[width=0.9\textwidth]{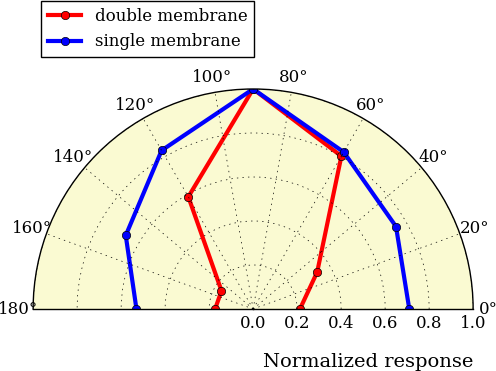}
   \end{center}
  \end{minipage}\\ 
  \begin{minipage}[]{0.495\textwidth}
    \begin{center}
      (a)
    \end{center}
  \end{minipage}
\hfill
  \begin{minipage}[]{0.495\textwidth}
    \begin{center}
      (b)
    \end{center}
  \end{minipage}
  \begin{center}
    \begin{minipage}[b]{\textwidth}
      \caption[]{\label{fig:directionality} Directionality of the response. In (a) the transfer function for a double sensor membrane is shown for various angles of the sensor with respect to the source. In (b) the resonance peak (normalized to the 90$^o$ response) is shown for two types of sensors as a function of the incident angle.}
  \end{minipage}
 \end{center}
\end{figure}

The experimental set-up allowed to change the orientation of the fiber hydrophone with respect to the source, so that the directionality could be determined. To do so, the transfer function has been determined at various orientations relative to the source for both types of fiber hydrophones. In figure \ref{fig:directionality} we show the transfer function for various angles with respect to the source for a double membrane sensor. As can be seen from the figure, the transfer function for the double membrane fiber hydrophone varies with the angle, while for the single membrane fiber hydrophone the effect is less pronounced. We furthermore observe that the direction sensitivity is not observed outside the resonance peak. The position of the resonance peak is independent of the angle of incidence.

\subsection{Pulse measurement}
\begin{figure}[ht]
   \begin{center}
   \includegraphics[width=0.5\textwidth]{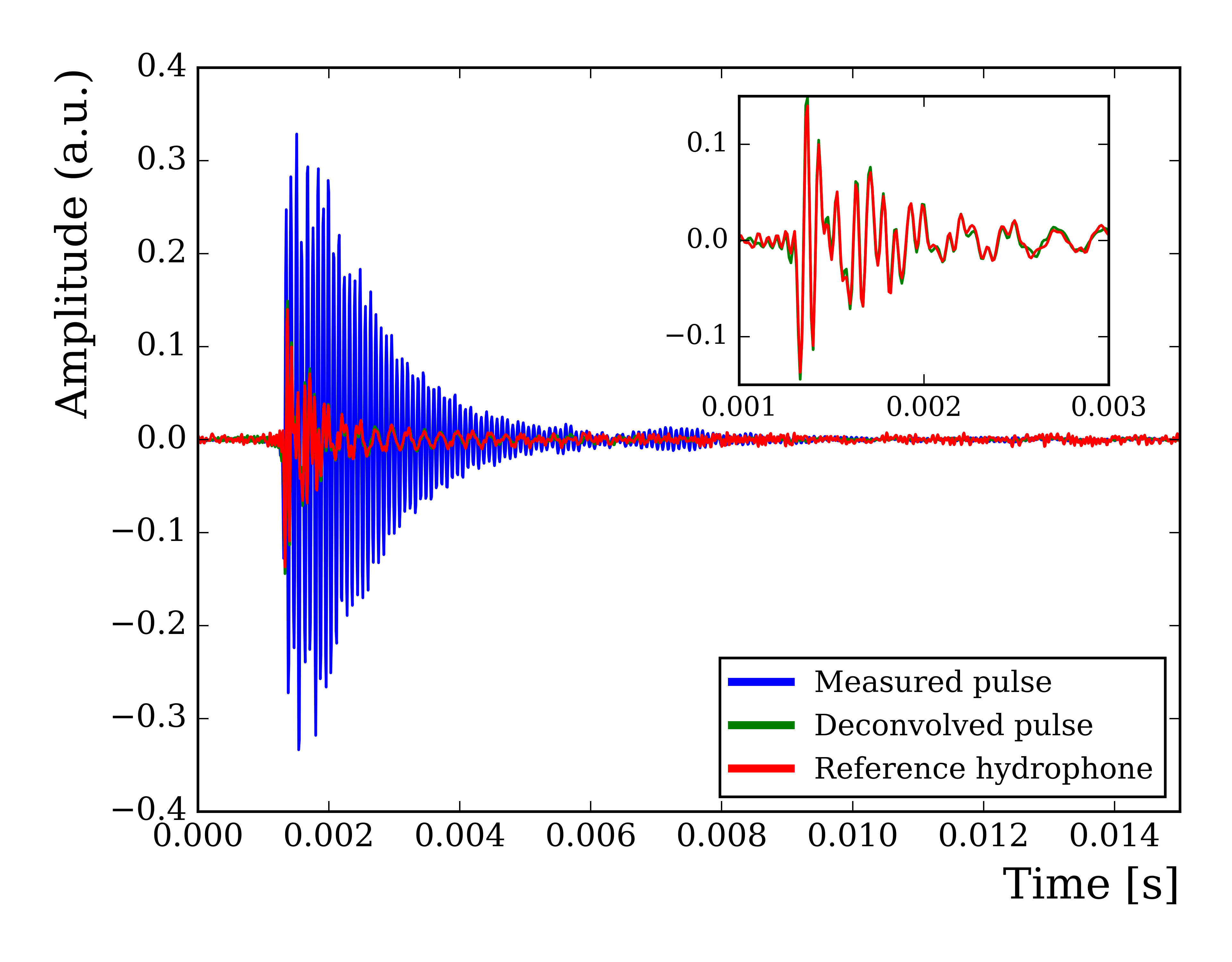}
   \end{center}
  \begin{center}
      \caption[]{\label{fig:pulse_deconvoluted} Measurement of a single pulse using both the fiber hydrophone and the reference hydrophone. In addition we show the deconvoluted pulse measured by the fiber hydrophone. }
 \end{center}
\end{figure}

The investigation of the (complex) transfer function of the fiber hydrophone is justified when we have a look at the single pulses or transients. A number of pulse trains have been recorded to this end. In figure \ref{fig:pulse_deconvoluted} we show an example of one such pulse, measured with both the fiber hydrophone and the reference hydrophone.

The mechanical eigenfrequency of the transducer is manifest as a resonant peak in the transfer function and as a significant amount of ringing in the time trace: a short transient signal shows a large tail that oscillates with the eigenfrequency. Knowing the transfer function of the hydrophone, we can deconvolute the transfer function to compare the resulting pulse with the reference hydrophone. To do so we take the transfer function as shown in figure \ref{fig:bodeplot} and do interpolation to match the transfer function to the amount of data points in the pulse. The result of the deconvolution is shown in the insert of figure \ref{fig:pulse_deconvoluted}. After deconvolution, both pulses are nearly indistinguishable, an indication that the transfer function function has been properly determined. Note that also the signal recorded by the reference hydrophone shows some ringing. Since the reference hydrophone has a flat frequency response (and hence no ringing is expected), the ringing is assumed to originate from the source. 

\subsection{Pressure stress tests}
\begin{figure}[ht]
   \begin{center}
   \includegraphics[width=0.75\textwidth]{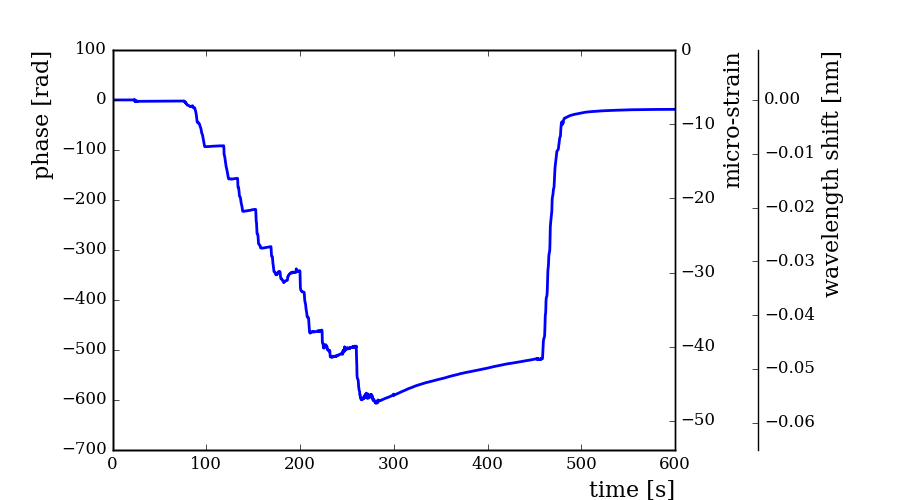}
   \end{center}
  \begin{center}
      \caption[]{\label{fig:pressure_cycle} Measurement of the phase change during a pressure cycle. The pressure increases in steps of roughly 20 bar up to a maximum of 140 bar.}
 \end{center}
\end{figure}

One of the challenging requirements to meet is the fact that the fiber hydrophones are operated in the high pressure environments of the deep sea. A first investigation to verify the compliance with the requirement has been carried out in a pressure stress test. To this end a fiber hydrophone was placed in a small hyperbaric pressure vessel and put under pressure of 140 bar. The pressure was applied at several steps. In figure \ref{fig:pressure_cycle} the time line of the measurement is shown: during a time period of 600 seconds the phase measurement in the interferometer has been monitored, while the ambient pressure increased by steps of roughly 20 bar. The pressure was applied manually (pump HP32-8 by Uraca) in steps of roughly 20 bar, while each step was followed by a few seconds of stable pressure until a maximum of 140 bar was reached. Then, the pressure was held to a constant level for about 3 minutes after which the pressure was released and returned to ambient pressure. The graph in figure \ref{fig:pressure_cycle} shows the wavelength shift during the complete pressure cycle. Note that the maximum pressure applied was limited by the fact that the vessel was put under pressure by hand.

The high pressure will slightly deform the transducer and the integrated fiber laser. This gives rise to a phase shift as measured in the interferometer in the interrogator and is plotted in figure \ref{fig:pressure_cycle}. Subsequently the wavelength shift of the fiber laser has been determined using equation \ref{eqn:dphase_dlambda}. The figure shows that due to the deformation of the transducer the shift of the central wavelength of the fiber laser will shift 60 pm at a pressure of 140 bar. Once the fiber hydrophone will be deployed to the deep sea, the central wavelength of the laser will remain at this value. Note that this measurement indicates that the pressure compensation does not remove the dependency of the wavelength of the fiber hydrophone on the applied depth completely, but enough for proper operation. 

\begin{figure}[ht]
   \begin{center}
   \includegraphics[width=0.65\textwidth]{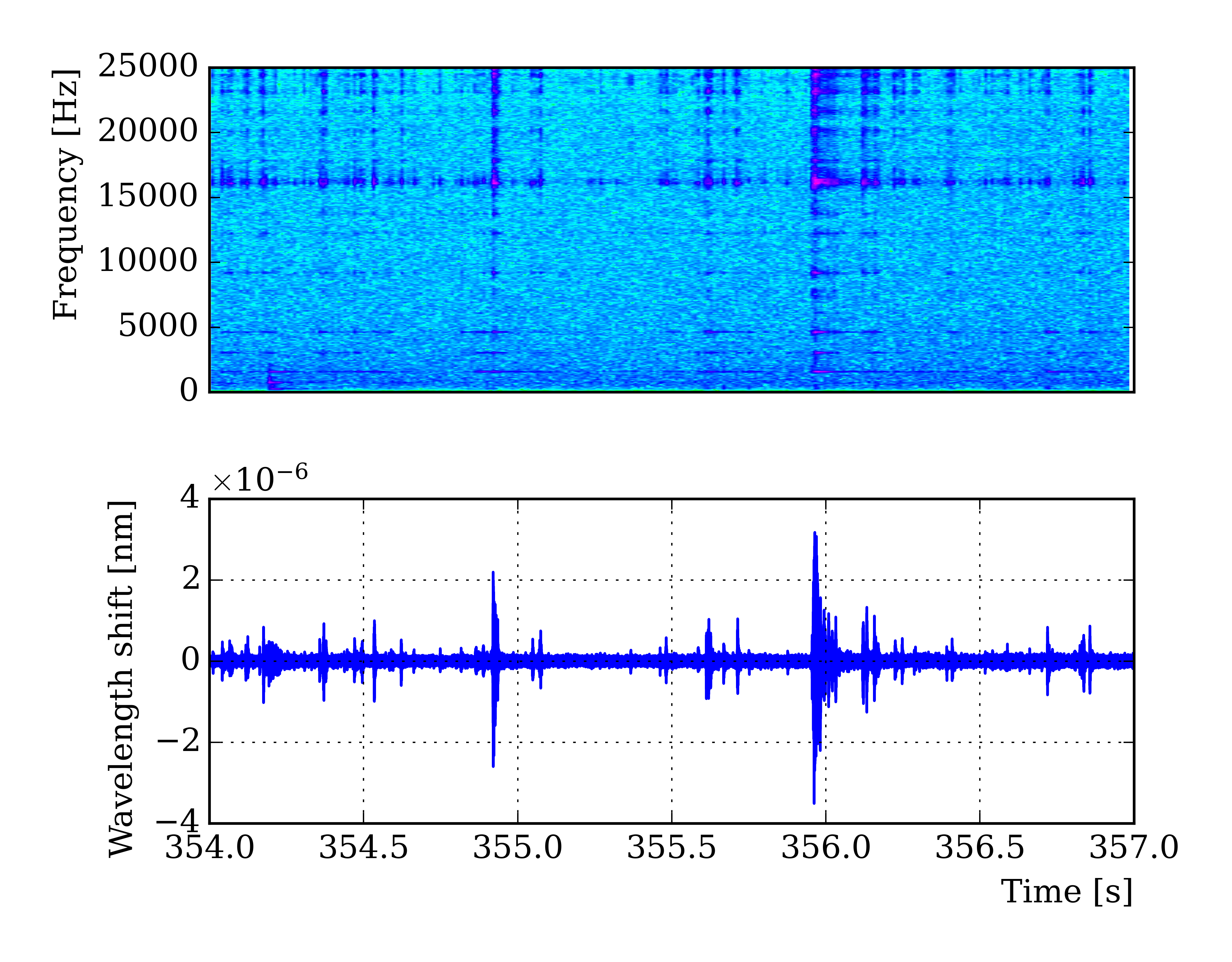}
   \end{center}
  \begin{center}
      \caption[]{\label{fig:spectrogram_354} Spectrogram (top) and time trace (bottom) of a brief event recorded during the pressure test.}
 \end{center}
\end{figure}

If we zoom in on the time trace of the measurement of the phase (and hence the wavelength) during the pressure qualification as shown in figure \ref{fig:pressure_cycle}, then we can plot again the spectrogram. The spectrogram of the period between 354 and 357 seconds after the start of the measurement is shown in the figure \ref{fig:spectrogram_354}. In the top panel we show the spectrogram, while in the bottom panel the time trace is plotted. The time trace is plotted after a high-pass filter (with a cut-off frequency of 500 Hz). The time trace shows spiky behavior: The measured wavelength does not only show a gradual decrease and subsequent increase during the measurement (as shown in figure \ref{fig:pressure_cycle}), but additional transient signals have been recorded. The origin of the signal is not clear; this could well be the deformation of the transducer, but also acoustic signals from the pressure vessel itself. We can use these transients to investigate one of the main features of the transfer function, i.e. the resonance peak. The spectrogram shows the position of the resonance peak at $\sim$16 kHz by the dark blue horizontal band. We can therefore conclude that the position of the resonance peak was not altered when the transducer was put under a static pressure of 140 bar. 

A measurement of the transfer function was carried out before and after the pressure cycle in the anechoic basin of which the results. There was no noticeable difference found between the two transfer functions: The mechanical resonance peak was found at the same frequency and the overall strain gauge factor remained identical. This indicates that the sensor and the attachment of the fiber to the transducer did not deform. In addition, the same directionality has been found as depicted in figure \ref{fig:directionality}. 

It should be noted that the pressure cycle in this experiment was much shorter than what can be expected during deployment in the applicable environment. If the fiber hydrophones were deployed in the deep sea, the rate of pressure change would be an order of magnitude smaller, resulting in a more gradual environmental transition.

\subsection{Sensitivity}
\begin{figure}[ht]
   \begin{center}
       \includegraphics[width=0.5\textwidth]{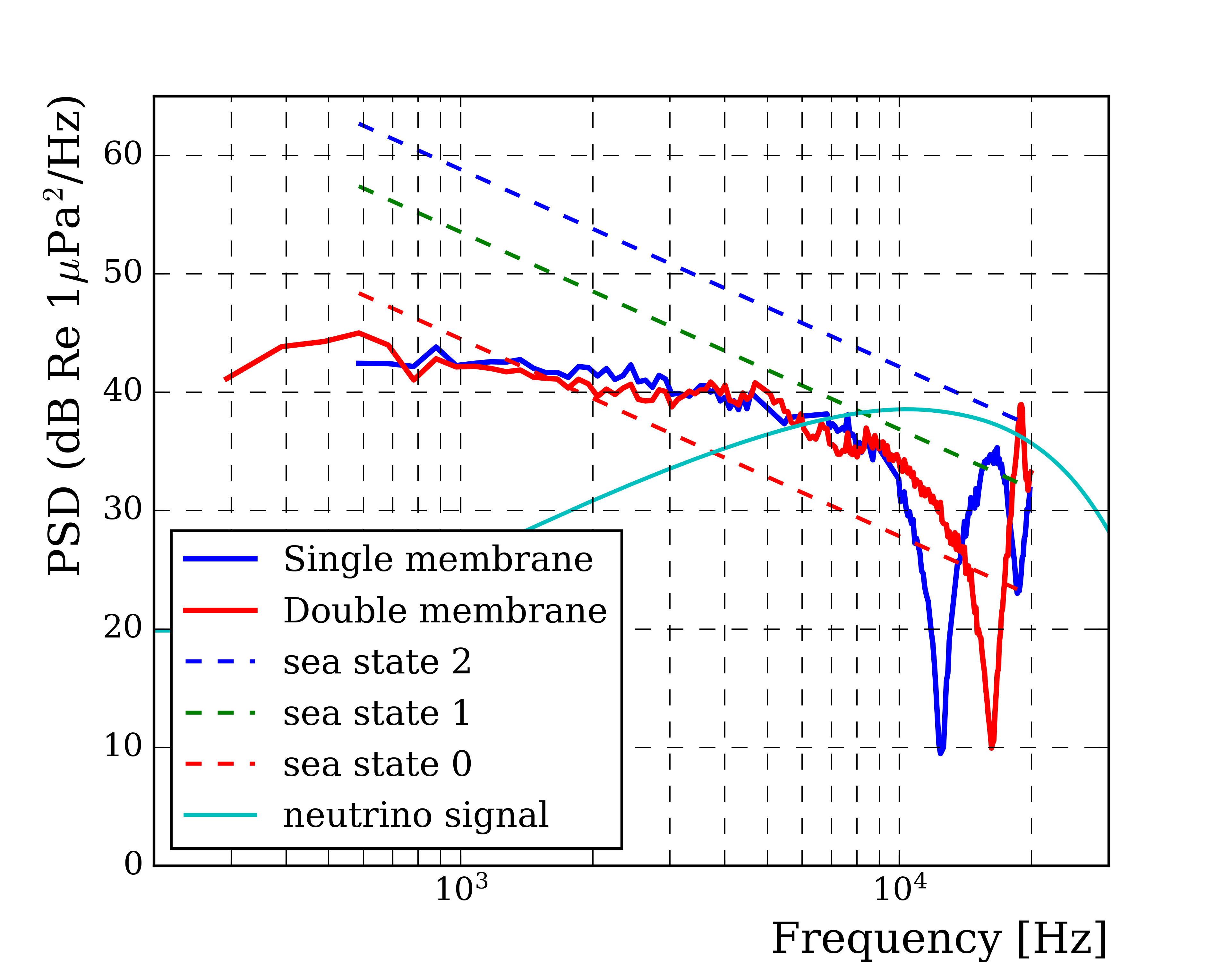}
   \end{center}
  \begin{center}
      \caption[]{\label{fig:sensitivity} Sensitivity of the two tested types of fiber hydrophones as a function of frequency. In addition ambient sea state noise levels are drawn for comparison as well as an example of a neutrino pulse. See text for details on sea state noise.}
 \end{center}
\end{figure}

Based on the measured transfer function and system noise, it is possible to determine the sensitivity of the fiber hydrophone. We therefore take the strain gauge factor as shown in figure \ref{fig:strainsensitivity_design} of the two types of transducers and the system noise as depicted in figure \ref{fig:RIN_fl}. The resulting sensitivity of the fiber hydrophone as a function of the acoustic frequency is plotted in figure \ref{fig:sensitivity}. It is illustrative to compare the sensitivity with the noise levels as expected in the deep sea. It is common to express the deep sea noise in terms of a sea state noise (\cite{Knudsen1948}), which runs from sea state 0 (SS0) to SS9. The expression of the sea states is provided in \cite{Ulrick1983, Short2005} and is given by:
\begin{equation}
    P(f, {\rm n}_s) = 10\log_{10}(f^{-\frac{5}{3}}) + 30\log_{10}({\rm n}_s + 1)   + 94.5\;\;\; \left[\rm{dB \; re} \;\frac{\mu Pa^2}{Hz} \right] \;\;\;,
\end{equation}
where $f$ is the frequency and n$_s$ the sea state. 

Some levels have been depicted in figure \ref{fig:sensitivity} as well and it can be concluded that the overall sensitivity compares to the lowest expected noise levels in the deep sea, varying between sea state 0 and 1. In particular, the sensitivity is enhanced down below sea state 0 at the position of the resonance peaks. Note further that the sensitivities of the two types of fiber hydrophones are comparable away from the resonance peaks. Furthermore, although not shown in the figure, the response of the fiber hydrophones is non-zero well above 20 kHz, which allows for higher frequency signals to be detected. Finally, as an example, the spectrum of a neutrino was added to the figure. The spectrum of the neutrino signal may vary with energy or distance and orientation with respect to the particle shower in the water. In this case a signal was simulation for a neutrino with energy of $10^{10}$ GeV at a distance of 100 m. 

\section{Conclusions}
\label{sec:concl}
This paper presents the development of a highly sensitive fiber hydrophone based on fiber laser technology, designed for potential use in future acoustic neutrino telescopes. To meet the demands of deep-sea operation, the sensor must function at ambient pressures of at least 10 MPa while detecting transient pressure signals in the range of several mPa. A pressure compensation mechanism has been implemented to offset static pressure levels.

The fiber hydrophone system has been subjected to a detailed investigation of the acoustic performance. A dedicated set up in an anechoic pool has been made for this purpose. Based on the presented results we arrive at the following conclusions. The transfer function of the fiber hydrophone follows the predictions of the finite element analysis that were carried out during the design phase: Two resonance peaks are found in the response of the fiber hydrophone: i) one Helmholtz resonance at lower frequencies ($<$ 500 Hz) and ii) one mechanical resonance of the transducer well above 10 kHz. The exact values of the eigenfrequencies differ from the predicted values, but the dynamical range of the sensors is found to be 200 Hz to 20 kHz, compliant with the main frequency components of a transient neutrino signal.

There is significant difference between the transfer functions of a single and a double membrane type fiber hydrophone. While the double membrane type shows an overall higher sensitivity, it also exhibits a substantial directionality, in particular around the resonance peaks. We converted the measured wavelength shift pressure sensitivity for a direct comparison with ambient sea state noise. It was found that the sensitivity of the technology is between sea state 0 and 1 for most of the dynamic range, which allows for detecting weak acoustic signals generated by neutrino interactions in the deep sea.

Measurements on pulses show the impact of a complex transfer function. As expected, transient signals exhibit significant ringing. However, after deconvolution of the fiber hydrophone response the pulse shape matches the reference pulse, confirming that the transfer function is well-characterized and can be corrected. As the pulse shape and size is directly related to the energy of the interacting neutrino, deconvolution is important. Various approaches can be employed, including the deconvolution method presented here, system identification techniques, or even convolution of a template bank with the fiber hydrophone response in the context of matched filtering. The discussion on how the deconvolution exactly is carried out once the fiber hydrophones are implemented in a full size neutrino telescope is beyond the scope of this paper.

Pressure qualification has been carried out to confirm the working principle of the static pressure compensation mechanism. The fiber hydrophone was placed in a vessel, after which a pressure of 140 bar (14 MPa) was applied. The fiber hydrophone showed no degradation in sensitivity after the pressure cycle. Transient signals that were observed during the pressure qualification showed that the mechanical resonance of the transducer did not change prior, during and post qualification. This may lead to the conclusion that the transfer function of the transducer under pressure is similar to the one outside the pressure vessel. Whether the overall sensitivity changes under pressure needs verification in a dedicated experiment. It requires an acoustic set-up in a high pressure environment. 

The concept of distributed sensing using optical fibers, combined with the results of the acoustic measurements and pressure qualification, makes this technology well-suited for implementation in a large acoustic neutrino telescope.

\section{Acknowledgment}
This research was funded by NWO, project nr 184.034.013.

\section{Bibliography}
\vspace{-0.5cm}
\bibliographystyle{unsrt} 
\bibliography{foh}

\begin{thebibliography}{10}

\bibitem{Askaryan1957}
G.A. Askaryan.
\newblock Hydrodynamic radiation from the tracks of ionizing particles in
  stable liquids.
\newblock {\em The Soviet Journal of Atomic Energy}, 3(2):921 -- 923, 1957.

\bibitem{Askaryan1979}
G.A. Askaryan, B.A. Dolgoshein, A.N. Kalinovsky, and N.V. Mokhov.
\newblock Acoustic detection of high energy particle showers in water.
\newblock {\em Nuclear Instruments and Methods}, 164(2):267--278, 1979.

\bibitem{Learned1979}
J.G. Learned.
\newblock Acoustic radiation by charged atomic particles in liquids: An
  analysis.
\newblock {\em Phys. Rev. D}, 19:3293--3307, Jun 1979.

\bibitem{Karg2006}
Timo Karg.
\newblock {\em {Detection of ultrahigh energy neutrinos with an underwater very
  large volume array of acoustic sensors: A simulation study}}.
\newblock PhD thesis, Erlangen - Nuremberg U., 2006.

\bibitem{Foster2005}
Scott Foster, Alexei Tikhomirov, Mark Milnes, John van Velzen, and Graham
  Hardy.
\newblock {A fiber laser hydrophone}.
\newblock In Marc Voet, Reinhardt Willsch, Wolfgang Ecke, Julian Jones, and
  Brian Culshaw, editors, {\em 17th International Conference on Optical Fibre
  Sensors}, volume 5855, pages 627 -- 630. International Society for Optics and
  Photonics, SPIE, 2005.

\bibitem{Goodman2008}
S.~{Goodman}, A.~{Tikhomirov}, and S.~{Foster}.
\newblock {Pressure compensated distributed feedback fibre laser hydrophone}.
\newblock In {\em Society of Photo-Optical Instrumentation Engineers (SPIE)
  Conference Series}, volume 7004 of {\em Society of Photo-Optical
  Instrumentation Engineers (SPIE) Conference Series}, June 2008.

\bibitem{Bevan2007}
S.~Bevan, S.~Danaher, J.~Perkin, S.~Ralph, C.~Rhodes, L.~Thompson, T.~Sloan,
  and D.~Waters.
\newblock Simulation of ultra high energy neutrino induced showers in ice and
  water.
\newblock {\em Astroparticle Physics}, 28(3):366--379, 2007.

\bibitem{Clara2020}
Clara~Gatius Oliver.
\newblock Acoustic signals from ultra high energy cosmic neutrinos.
\newblock Master's thesis, University of Amsterdam, 2020.

\bibitem{Buis2013}
E.~J. Buis, E.~J.~J. Doppenberg, R.~A. Nieuwland, and P.~M. Toet.
\newblock {Fibre laser hydrophones for cosmic ray particle detection}.
\newblock {\em JINST}, 9:C03051, 2014.

\bibitem{Buis2017}
Ernst-Jan Buis, Ed~Doppenberg, Robert Lahmann, and Peter Toet.
\newblock {A large fiber sensor network for an acoustic neutrino telescope}.
\newblock {\em EPJ Web Conf.}, 135:06006, 2017.

\bibitem{Cranch2008}
Geoffrey~A. Cranch, Gordon M.~H. Flockhart, and Clay~K. Kirkendall.
\newblock Distributed feedback fiber laser strain sensors.
\newblock {\em IEEE Sensors Journal}, 8(7):1161--1172, 2008.

\bibitem{Foster2017}
Scott~B. Foster, Geoffrey~A. Cranch, Joanne Harrison, Alexei~E. Tikhomirov, and
  Gary~A. Miller.
\newblock Distributed feedback fiber laser strain sensor technology.
\newblock {\em Journal of Lightwave Technology}, 35(16):3514--3530, 2017.

\bibitem{Kersey_1996}
A.D. Kersey.
\newblock A review of recent developments in fiber optic sensor technology.
\newblock {\em Optical Fiber Technology}, 2(3):291--317, 1996.

\bibitem{Bertholds88}
A.~Bertholds and R.~Dandliker.
\newblock Determination of the individual strain-optic coefficients in
  single-mode optical fibres.
\newblock {\em Journal of Lightwave Technology}, 6(1):17--20, 1988.

\bibitem{Knudsen1948}
V.O Knudsen, R.S. Alford, and J.W. Emling.
\newblock Underwater ambient noise.
\newblock {\em Journal of Marine Research}, 7(2):410--429, 1948.

\bibitem{Ulrick1983}
R.J. Urick.
\newblock {\em Principles of Underwater Sound}.
\newblock McGraw-Hill, 1983.

\bibitem{Short2005}
J.R. Short.
\newblock High-frequency ambient noise and its impact on underwater tracking
  ranges.
\newblock {\em IEEE Journal of Oceanic Engineering}, 30(2):267--274, 2005.

\end{thebibliography}

\end{document}